\documentclass[aps,pra,twocolumn,floatfix,superscriptaddress,showpacs]{revtex4}
\usepackage{amssymb}
\usepackage{latexsym}
\usepackage{amsmath}
\usepackage{graphicx}
\usepackage{float}
\usepackage[applemac]{inputenc}
\usepackage[nooneline]{subfigure}

\begin{document}

\title{Magnetic properties of a strongly correlated system on the Bethe lattice}

\author{F. P. Mancini}

\affiliation{Dipartimento di Fisica {\it ``E. R. Caianiello"} and
CNR-SPIN, UOS di Salerno, \\
 Universit\`a degli Studi di Salerno,
Via Ponte don Melillo I-84084 Fisciano (SA), Italy}


 \affiliation{I.N.F.N. Sezione di Perugia, Via A. Pascoli,
I-06123 Perugia, Italy}



\begin{abstract}
We study the influence of an external magnetic field $h$ on the
phase diagram of a system of Fermi particles living on the sites
of a Bethe lattice with coordination number $z$ and interacting
through on-site $U$ and nearest-neighbor $V$ interactions. This is
a physical realization of the extended Hubbard model in the
narrow-band limit. Our results establish that the magnetic field
may dramatically affect  the critical temperature below which a
long-range charge ordered phase is observed, as well as the
behavior of physical quantities, inducing, for instance,
magnetization plateaus in the magnetization curves. Relevant
thermodynamic quantities - such as the specific heat and the susceptibility -
are also investigated at finite temperature by varying the on-site
potential, the particle density and the magnetic field.
\end{abstract}


\maketitle


Statistical models on the Bethe lattice are of considerable
interest since they admit a direct analytical approach for a
number of problems that may be otherwise intractable on Euclidean
lattices. Due to the peculiar structure of the lattice, several
interesting physical problems involving interactions are exactly
solvable when defined on the Bethe lattice \cite{baxter}. There is
a general interest in the study of models defined on such lattices
which goes beyond physics. Being a dendrimer of infinite
generation, the Bethe lattice is attractive both for basic and
applied interdisciplinary research involving chemistry, physics,
biology, pharmaceutics, and medicine \cite{dendri}.

In a recent paper, we provided a comprehensive and systematic
analysis of the extended Hubbard model in the narrow-band (atomic)
limit (AL-EHM) on the Bethe lattice with arbitrary coordination
number $z$ \cite{mancinis09}. Within the Green's function and
equations of motion formalism, we exactly solved the model and, by
considering relevant physical quantities in the whole space of the
model parameters, we investigated the finite temperature phase
diagram, for both attractive and repulsive on-site and intersite
interactions. The aim of the present paper is twofold. First, we
would like to further develop our previous work, by extending it
to a more general situation in which a finite magnetization may be
induced by an external magnetic field. Secondly, the AL-EHM on the
Bethe lattice exhibits interesting features at low temperatures
such as the existence of magnetic plateaus. Here, we address the
problem of determining the effect on the phase diagram and on the
behavior of several thermodynamic quantities of the presence of a
uniform magnetic field, introduced through a Zeeman term.  We
study the properties of the system as functions of the external
parameters $n$, $T/V$, $U/V$ and $h/V$, allowing for the on-site
interaction $U$ to be both repulsive and attractive. In fact, the
parameter $U$ can represent the effective interaction coupling
taking into account also other interactions. Throughout the paper,
we consider a repulsive intersite interaction $V$  and we set
$V=1$ as the unit of energy, taking the Boltzmann's constant
$k_B =1$. In the absence of a magnetic field,
the phase diagram in the plane ($n,T$) exhibits a transition line along the temperature
axis, below which translational invariance is broken \cite{mancinis09}. The Bethe
lattice effectively splits in two sublattices with different
thermodynamic properties. As a result, a charge ordered (CO)
phase, characterized by a different distribution of the electrons
in alternating shells, is established for pertinent values of the
particle density. A finite magnetic field can have dramatic effect
on the phase diagram: for instance, when $U<0$, the reentrant
behavior observed for $h=0$  \cite{mancinis09} disappears and one finds, above a
certain value of $h$, a CO phase for $n>1/z$. This should be compared
with the case $ h=0 $, where the CO phase is observed only for $ n>2/z$ when
$ U<0 $ \cite{mancinis09}. The characteristic
lobe structure, exhibited by the critical temperature as a function
of the particle density, shrinks by augmenting the magnetic field.
By further increasing $h$, the CO phase is suppressed in the particle density range
$0.75<n<0.85$ and $1.15<n<1.25$ at $h = h_T=z V/2- U/2$; the lobe
splits in three separated lobes centered around $n=0.5$, 1, and
$1.5$, respectively. The central lobe eventually vanishes by
increasing $h$. A similar behavior is observed also for $U>0$.

The magnetic properties of the system depend on the value of the
particle density and of the on-site potential. At low
temperatures, and for attractive on-site interactions, the
magnetic field does not play any role if its intensity is $h<\vert
U \vert/2$: the ground state is a collection of shells with doubly
occupied sites (doublons) surrounded by empty shells. The magnetic
energy is not strong enough to break the doublons. On the other
hand, for strong repulsive on-site interactions, it is sufficient
a small nonzero value of the magnetic field to have a finite
magnetization. In the intermediate region, the competition among
$U$, $V$ and $h$ determines the phase structure. For all values of
the particle density, one finds a critical value  of the magnetic field $h_s$ - dramatically
depending on $U$ and $n$ - above which
the ground state is paramagnetic. In this state, every occupied
site contains one and only one electron, aligned along the
direction of $h$. For strong repulsive on-site interactions,
$h_s=0$, i.e., the spin are polarized as soon as the magnetic
field is turned on. Furthermore, for attractive on-site
interactions and $0.5<n \le 1$, one observes the existence of two
critical fields, namely: $h_c$, up to which no magnetization is
observed, and $h_s$, marking the beginning of full polarization.
This is analogous to the finite field behavior of the $S=1$
Haldane chain \cite{haldane}.

The addition of a homogeneous magnetic field does not dramatically
modify the framework of calculation given in Ref.
\cite{mancinis09}, provided one takes into account the breakdown
of the spin rotational invariance. For the sake of
comprehensiveness, in the next section, we briefly report the
analysis leading to the exact solution of the AL-EHM on the Bethe
lattice in the presence of a magnetic field. Then, we investigate
the phase diagram in the space $n$, $T/V$, $U/V$ by varying $h$
and we observe that, above a critical value of $h$, the CO region
shrinks due to the presence of a finite magnetic field. We also
analyze the magnetic properties of the system and find magnetic
plateaus in the magnetization curve. Finally, last section is
devoted to our conclusions.

\section{Exactly solvable model}
\label{sec_II}

When defined on the Bethe lattice with coordination number $z$,
the narrow-band limit of the extended Hubbard model, in the
presence of an external homogeneous magnetic field,  can be
described by the following Hamiltonian:
\begin{equation}
\label{hamilt}
H=-\mu n(0)+UD(0)-h \, n_3(0)+\sum_{p=1}^z H^{(p)}.
\end{equation}
$H^{(p)}$ is the Hamiltonian of the $p$-th sub-tree rooted at the
central site $(0)$ and can be written as
\begin{equation}
\label{bethe_H} H^{(p)}=-\mu \, n(p)+UD(p)-h \,
n_3(p)+Vn(0)n(p)+\sum_{m=1}^{z-1} H^{(p,m)}.
\end{equation}
Here, ($p$) ($p=1,\ldots z)$ are the nearest-neighbor sites of
$(0)$, also termed the first shell. $H^{(p,m)}$ describes the
$m$-th sub-tree rooted at the site ($p$):
\begin{equation}
\label{eq6}
\begin{split}
H^{(p,m)} &=-\mu \, n(p,m)+U\, D(p,m)-h \,
n_3(p,m)\\
&+V\,n(p)\,
n(p,m)+\sum_{q=1}^{z-1} H^{(p,m,q)} .
\end{split}
\end{equation}
$(p,m)$ ($m=1,\ldots
z-1)$ and $(0)$ are the nearest-neighbors of the site ($p$). $H^{(p,m,q)}$ is the Hamiltonian of the $q$-th
sub-tree rooted at the site $(p,m)$. The
process may be continued indefinitely.
$U$ and $V$ are the strengths of the local and intersite
interactions, respectively; $\mu$ is the chemical potential,
$n(i)=n_{\uparrow
}(i)+n_{\downarrow }(i)$ and $D(i)=n_{\uparrow }(i)n_{\downarrow
}(i)=n\left( i\right) \left[ n\left( i\right) -1\right] /2$ are
the charge density and double occupancy operators at site
 $\bf i$, respectively. $n_3 (i)$ is the third component of
the spin density operator, also called the electronic Zeeman term,
\begin{equation}
\label{eq2}
n_3 (i)=n_\uparrow (i)-n_\downarrow (i)=c_\uparrow
^\dag (i)c_\uparrow (i)-c_\downarrow ^\dag (i)c_\downarrow (i).
\end{equation}
Here we do not consider the orbital interaction with the magnetic
field. As usual, $n_{\sigma }(i)=c_{\sigma }^{\dag }(i)c_{\sigma
}(i)$ with $\sigma =\left\{ {\uparrow ,\downarrow }\right\}$,
where $c_{\sigma }(i)$ ($c_{\sigma }^{\dag }(i))$ is the fermionic
annihilation (creation) operator of an electron of spin $\sigma $
at site $\bf{i}$, satisfying canonical anticommutation relations.
We use the Heisenberg picture: $i=\left( {\bf i}, t\right)$, where
$\bf{i}$ stands for the lattice vector $\bf{R}_{i}$.

The exact solution of the model can be obtained by using the
equations of motion approach in the context of the composite
operator method \cite{manciniavella}, which is based on the choice
of a convenient operatorial basis. For our purposes, the suitable
field operators are the Hubbard operators, $\xi_{\sigma}
(i)=[1-n(i)]c_{\sigma}(i)$ and $\eta_{\sigma}
(i)=n(i)c_{\sigma}(i)$, which satisfy the equations of motion:
\begin{equation}
\begin{split}
i\frac{\partial }{\partial t}\xi _\sigma (i)&=-(\mu +\sigma h)\xi
_\sigma
(i)+zV\xi _\sigma (i)n^\alpha (i) \\
 i\frac{\partial }{\partial t}\eta _\sigma (i)&=-(\mu -U+\sigma h)\eta
_\sigma (i)+zV\eta _\sigma (i)n^\alpha (i).
\end{split}
\label{eq4a}
\end{equation}
Hereafter, for a generic operator $\Phi \left( i\right) $ we shall
use the notation $\Phi ^{\alpha }(i)=\sum\nolimits_{p=1}^{z}\Phi
(i,p)/z$, $ (i,p)$ being the first nearest-neighbors of the site
${\bf i}$. The Heisenberg equations \eqref{eq4a} contain the
higher-order nonlocal operators $\xi_{\sigma} (i)n^{\alpha }(i)$
and $\eta_{\sigma} (i)n^{\alpha }(i)$. By taking time derivatives
of the latter, higher-order operators are generated. This process
may be continued and an infinite hierarchy of field operators is
created. However, since the number $n(i)$ and the double occupancy
$D(i)$ operators satisfy the following algebra
\begin{equation}
\begin{split}
n^{p}\left( i\right) &=n\left( i\right) +a_{p}D(i),  \\
D^{p}\left( i\right) &=D(i), \\ n^{p}\left( i\right)
D(i)&=2D(i)+a_{p}D(i),
\end{split}
 \label{al_prop}
\end{equation}
where $p  \geq 1$ and $a_{p}=2^{p}-2$, it is straightforward to
establish the following recursion rule \cite{Mancini05}:
\begin{equation}
\lbrack n^{\alpha }(i)]^{k}=\sum_{m=1}^{2z}A_{m}^{(k)}[n^{\alpha
}(i)]^{m},
 \label{eq5}
\end{equation}
which allows one to write the higher-power expressions of the
operator $n^\alpha (i)$ in terms of the first $2z$ powers. The
coefficients $A_{m}^{(k)}$ are rational numbers, satisfying the
relations $\sum_{m=1}^{2z}A_{m}^{(k)}=1$ and $A_{m}^{(k)}=\delta
_{m,k}$ $(k=1,\ldots ,2z)$ \cite{mancini2,mancini05b}. The
recursion relation \eqref{eq5} allows one to close the hierarchy
of equations of motion. As a result, a complete set of
eigenoperators of the Hamiltonian \eqref{hamilt} can be found.
To this end, one  defines the composite field operator
\begin{equation}
\label{eq8} \psi (i)=
\begin{pmatrix}
\psi^{(\xi )}(i)  \\
\psi^{(\eta )}(i)
\end{pmatrix}
=
\begin{pmatrix}
\psi_{\uparrow} ^{(\xi )}(i)  \\
\psi_{\downarrow} ^{(\xi )}(i)  \\
\psi_{\uparrow} ^{(\eta )}(i)\\
\psi_{\downarrow} ^{(\eta )}(i)
\end{pmatrix},
\end{equation}
where
\begin{equation}
\label{eq10}
\begin{split}
 \psi_\sigma ^{(\xi )}(i)&=
\begin{pmatrix}
 \xi_\sigma (i)  \\
 \xi_\sigma (i)[n^\alpha (i)]  \\
 \vdots   \\
\xi_\sigma (i)[n^\alpha (i)]^{2z}
\end{pmatrix},
\\
 \psi_\sigma^{(\eta )}(i) &=
\begin{pmatrix}
 \eta_\sigma (i)  \\
 \eta_\sigma (i)[n^\alpha (i)]  \\
 \vdots   \\
\eta_\sigma (i)[n^\alpha (i)]^{2z}
\end{pmatrix}.
\end{split}
\end{equation}
With respect to the case of zero magnetic field \cite{mancinis09},
the degrees of freedom have doubled, since one has taken into
account the two nonequivalent directions of the spin. By
exploiting the algebraic properties of the operators $n(i)$ and
$D(i)$, and the recursion rule \eqref{eq5}, it is easy to show
that the fields $\psi ^{(\xi )}(i)$ and $\psi ^{(\eta )}(i)$ are
eigenoperators of the Hamiltonian \eqref{hamilt}
\cite{mancini05b}:
\begin{equation}
\label{eq14}
\begin{split}
 i\frac{\partial }{\partial t}\psi ^{(\xi )}(i)&=[\psi ^{(\xi
)}(i),H]=\varepsilon ^{(\xi )}\psi ^{(\xi )}(i) ,\\
 i\frac{\partial }{\partial t}\psi ^{(\eta )}(i)&=[\psi ^{(\eta
)}(i),H]=\varepsilon ^{(\eta )}\psi ^{(\eta )}(i).
 \end{split}
\end{equation}
$\varepsilon ^{(\xi )}$ and $\varepsilon ^{(\eta )}$ are the
  energy matrices:
\begin{equation}
\label{eq15}
\varepsilon ^{(\xi )}=\left( {{\begin{array}{*{20}c}
 {\varepsilon _\uparrow ^{(\xi )} }  & 0  \\
 0  & {\varepsilon _\downarrow ^{(\xi )} }  \\
\end{array} }} \right),
\qquad  \varepsilon ^{(\eta )}=\left( {{\begin{array}{*{20}c}
 {\varepsilon _\uparrow ^{(\eta )} }  & 0  \\
 0  & {\varepsilon _\downarrow ^{(\eta )} }  \\
\end{array} }} \right),
\end{equation}
where $\varepsilon _\sigma ^{(\xi )}$ and $\varepsilon_\sigma
^{(\eta)}$ are the $(2z+1)\times (2z+1)$ matrices given by
\begin{widetext}
\begin{equation}
\label{eq16}
 \varepsilon _\sigma ^{(\xi )} =
\begin{pmatrix}
  -\mu -\sigma h  &   zV  &   0 &    \ldots &   0 &   0  &   0  \\
  0  &   -\mu -\sigma h  &   zV &    \ldots &   0 &   0  &    0  \\
  0  &   0  &   -\mu -\sigma h  &    \ldots &   0 &   0  &   0  \\
  \vdots  &   \vdots  &   \vdots  &    \vdots &   \vdots  &   \vdots  &   \vdots  \\
  0  &   0 &   0 &   \ldots  &   -\mu -\sigma h &   zV  &   0  \\
  0  &   0  &    0  &   \ldots &   0  &  -\mu -\sigma h  &   zV  \\
  0  &   zVA_1^{2z+1}  &
zVA_2^{2z+1} &   \ldots  &
zVA_{2z-2}^{2z+1} &   zVA_{2z-1}^{2z+1} &
-\mu-\sigma h+zVA_{2z}^{2z+1}
\end{pmatrix},
\end{equation}
\begin{equation}
\label{eq17} \varepsilon _\sigma ^{(\eta )} =
\begin{pmatrix}
   U-\mu -\sigma h  &   zV  &   0 &    \ldots &   0 &   0  &   0  \\
  0  &   U-\mu -\sigma h  &   zV &    \ldots &   0 &   0  &    0  \\
  0  &   0  &   U-\mu -\sigma h  &    \ldots &   0 &   0  &   0  \\
  \vdots  &   \vdots  &   \vdots  &    \vdots &   \vdots  &   \vdots  &   \vdots  \\
  0  &   0 &   0 &   \ldots  &   U-\mu -\sigma h &   zV  &   0  \\
  0  &   0  &    0  &   \ldots &   0  &  U-\mu -\sigma h  &   zV  \\
  0  &   zVA_1^{2z+1}  &
zVA_2^{2z+1} &   \ldots  &
zVA_{2z-2}^{2z+1} &   zVA_{2z-1}^{2z+1} &
U-\mu-\sigma h+zVA_{2z}^{2z+1}
\end{pmatrix}.
\end{equation}
\end{widetext}
The eigenvalues of the matrices $\varepsilon_{\sigma} ^{(\xi )}$
and $\varepsilon_{\sigma} ^{(\eta )}$ are
\begin{equation}
\label{EHM_7}
\begin{split}
 E^{(\xi )}_{p,\sigma}&=-\mu -\sigma h +(p-1)V,     \\
 E_{p,\sigma}^{(\eta )}&=U-\mu -\sigma h +(p-1)V ,
\end{split}
\end{equation}
with $p=1,\ldots ,2z+1$. The Hamiltonian has now been formally
solved since, for any coordination number of the underlying Bethe
lattice, one has found a closed set of eigenoperators and
eigenenergies. As a result, one may solve the model and compute
observable quantities.

Since the addition of a homogeneous magnetic field does not
dramatically modify the framework of calculation given in Ref.
\cite{mancinis09}, here we report only some details of the
calculations and refer the interested reader to Ref.
\cite{mancinis09} for a comprehensive analysis. Upon splitting the
Hamiltonian \eqref{hamilt} as
\begin{equation}
\label{eq8213}
\begin{split}
 H&=H_0^{(i)} +H_I^{(i)} ,\\
 H_I^{(i)} &=zVn(i)n^\alpha (i),
 \end{split}
\end{equation}
it is immediate to notice that, with respect to the case $h=0$,
only $H_0^{(i)}$ is modified by the presence of the magnetic field
$h$. Therefore, all the calculations relative to $H_I^{(i)}$
are not modified; the changes induced by the presence of $h$
will concern only the calculations involving $H_0$. In
particular, the statistical average of any operator $O$ can be
expressed as
\begin{equation}
\label{EHM_18}
 \langle O\rangle =\frac{\langle Oe^{-\beta H_I^{(i)}
}\rangle _{0,\bf i}} {\langle e^{-\beta H_I^{(i)} }\rangle_{0, \bf
i} }.
\end{equation}
The symbol $\langle \cdots \rangle_{0, \bf i}$ stands for the
thermal average with respect to the reduced Hamiltonian
$H_0^{(i)}$: i.e., $\langle \cdots \rangle_{0, \bf i} =Tr\{\cdots
e^{-\beta H_0^{(i)}}\}/Tr\{e^{-\beta H_0^{(i)}}\}$. Equation
\eqref{EHM_18} allows us to express the thermal averages with
respect to the complete Hamiltonian $H$ in terms of thermal
averages with respect to the reduced Hamiltonian $H_0$, which
describes a system where the original lattice has been reduced to
the site ${\bf i}$ and to $z$ unconnected sublattices. As a
consequence, in the $H_0$-representation, correlation functions
connecting sites belonging to disconnected sublattices can be
decoupled. Within this scheme, all the local correlators necessary
to compute the Green's functions \cite{mancinis09} can be written
as functions of the parameters
\begin{equation}
\begin{split}
X_i &= \langle n^{\alpha}(i)\rangle_{0,\bf i}
=\frac{1}{z}\sum_{p=1}^z \langle n(i_p )\rangle_{0,\bf i} , \\
Y_i &= \langle D^{\alpha}(i) \rangle_{0,\bf i} =
\frac{1}{z}\sum_{p=1}^z \langle D(i_p )\rangle_{0,\bf i} ,
\end{split}
\end{equation}
in terms of which one may find a solution of the model. In the
above equation, $i_p$ ($p=1,\ldots,z $) is an arbitrary
neighboring site of $\bf i$.

A repulsive intersite interaction disfavors the occupation of
neighboring sites. At low temperatures, this may lead to a CO
phase characterized by a nonhomogeneous distribution of the
electrons in alternating shells \cite{mancinis09}. In order to
capture this phase, we shall divide the lattice into two
sublattices: $A$ contains the central point ($0$) and the even
shells, the sublattice $B$ contains the odd shells. Then, one
requires the following boundary condition to hold:
\begin{subequations}
\begin{equation}
\label{eq101} \langle n(i) \rangle =\left\{ {{\begin{array}{*{20}c}
 {n_A } \\
 {n_B }  \\
\end{array} }} \right.\quad {\begin{array}{*{20}c}
 {i\in A} ,\\
 {i\in B}, \\
\end{array} }
\end{equation}
\begin{equation}
\label{eq102}
n=\frac{1}{N} \sum_i \langle n(i)\rangle= \frac{1}{2}(n_{A}+n_B).
\end{equation}
\label{boundcond}
\end{subequations}
Let us take two distinct sites $i\in A$ and $j\in B$. We require
that the expectation values of the particle density and of the
double occupancy operators at the site $i$ are equal to the ones
of the neighboring sites of $j$ and viceversa. As a consequence,
the number of unknown correlators is four: the parameters $X_i$,
$X_j$, $Y_i$, $Y_j$ are determined by the equations
\begin{equation}
\label{eq8225}
\begin{split}
 \langle n(i)\rangle &=\langle n(j_p )\rangle    \\
 \langle D(i)\rangle &=\langle D(j_p )\rangle
 \end{split}
  \quad \quad
\begin{split}
\langle n(i_p )\rangle &=\langle n(j)\rangle,
\\
\langle D(i_p )\rangle &=\langle
 D(j)\rangle .
 \end{split}
\end{equation}

After lengthy but straightforward calculations,  one can find
analytical expressions for the average values of the particle
density and of the double occupation  operators in terms of the
parameters $X_i$ and $Y_i$, namely:
\begin{equation}
\begin{split}
\label{eq_nD_1}
 \langle {n(i)}\rangle &=\frac{f(1+k^2)\,F_i ^z+2gk \,G_i ^z}{k+f(1+k^2)\,F_i ^z+ gk\,G_i^z},
 \\
  \langle {D(i)} \rangle &=\frac{gk\,G_i
^z}{k+f(1+k^2)\,F_i ^z+ gk\,G_i^z},
 \end{split}
\end{equation}
 and
\begin{equation}
\label{eq_nD_2}
\begin{split}
\langle {n(i_p )} \rangle &=
\frac{1}{\cal F}
\left[ k X_i +f(1+k^2) \,K \left( {X_i
+2aY_i } \right)F_i ^{z-1}
\right. \\
&+ \left.k g\,K^2 \left( {X_i +2d\,Y_i }
\right) G_i^{z-1} \right], \\
 \langle {D(i_p )} \rangle
&=\frac{k Y_i}{\cal F} \left[1+f(1+k^2) K^2F_i ^{z-1}+gk\,K^4 G_i
^{z-1}\right].
 \end{split}
\end{equation}
where $f=e^{\beta \mu}$,  $g=e^{\beta (2\mu-U)}$, and $k=e^{\beta
h}$, and we used the definitions
\begin{equation}
\label{eq8227}
\begin{split}
 F_i &=1+aX_i +a^2Y_i , \\
 G_i &=1+dX_i +d^2Y_i ,\\
\cal F &=k+f(1+k^2)\,F_i ^z+k g\,G_i ^z,
\end{split}
\end{equation}
with $K=e^{-\beta V}$, $a= K-1$, and $d=K^2-1$. Similarly, also
the third component of the spin density can be written as a
function of the parameters $X_i$ and $Y_i$:
\[
 \langle {n_3 (i)}\rangle =\frac{f(1-k^2)\,F_i ^z}{k+f(1+k^2)\,F_i ^z+ gk\,G_i^z},
\]
It is not difficult to show that the magnetization $m=\langle n_3 (i)\rangle$ can
be written in terms of the particle density and double occupancy as:
$\langle n_3 (i)\rangle =\tanh (\beta h)[n(i)-2D(i)]$.
Equations \eqref{eq8225}, together with Eq. \eqref{eq102} - which
fixes the chemical potential $\mu$ - constitute a system of
coupled equations allowing us to ascertain the five parameters
$\mu$, $X_A$, $X_B$, $Y_A$, and $Y_B$ in terms of the external
parameters of the model, namely: $n$, $h$, $U$, $V$, and $T$. Once these
quantities are known, all the properties of the model can be
computed.

\section{Phase diagram and magnetic properties}

\begin{figure}[t]
\centerline{\includegraphics[scale=0.31]{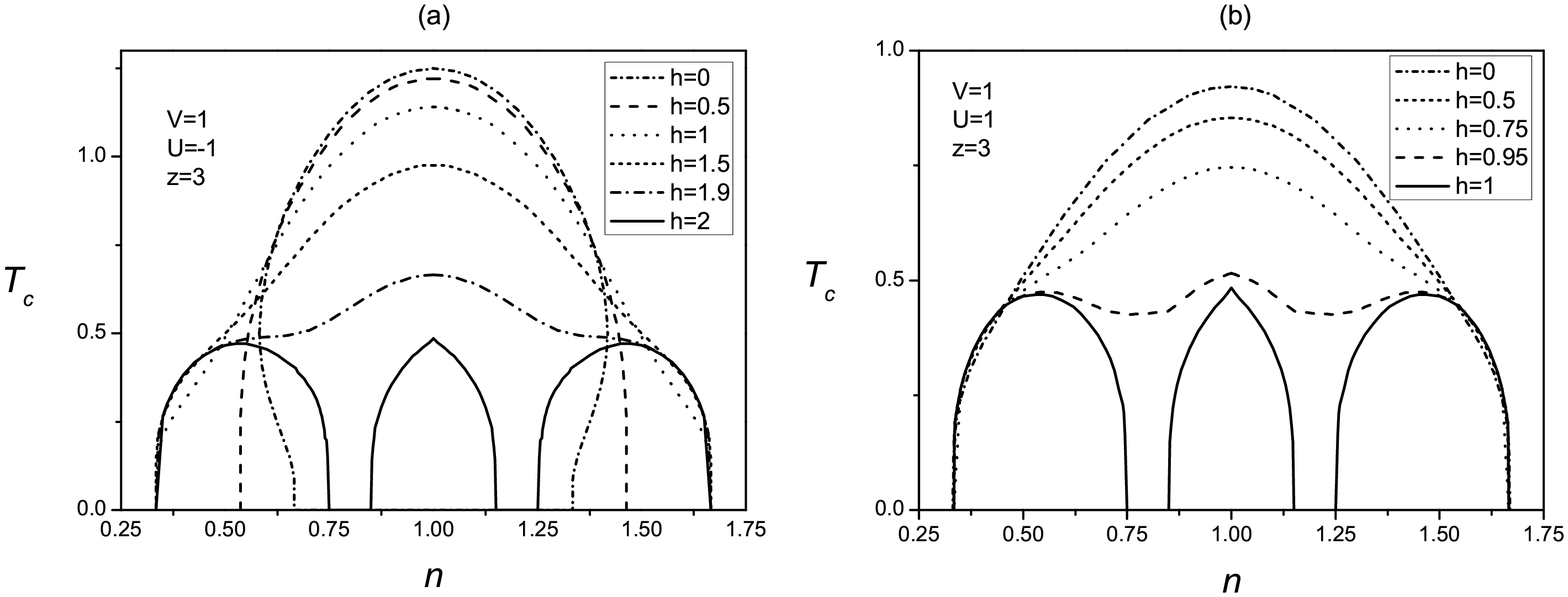}}
\caption{\label{pha_diam} Phase diagram in the plane ($T,n$) for
$V=1$ and $z=3$ and several values of $h$, and for (a) $U=-1$; (b)
$U=1$.}
\end{figure}

In this section, we derive the phase diagram by numerically
solving the set of equations \eqref{eq8225}. In the absence of a
magnetic field, we find regions of the $(U,n,T)$ 3D space
characterized by a spontaneous breakdown of translational
invariance \cite{mancinis09}. In these regions, the population of the two
sublattices $A$ and $B$ is not equivalent: the system has entered
a finite temperature long-range CO phase. Upon decreasing the
temperature, the distribution of the electrons becomes more
inhomogeneous. In the presence of a magnetic field, the phase
structure is determined by the three competing terms of the
Hamiltonian: the repulsive intersite potential (disfavoring the
occupation of neighboring sites), the magnetic field (aligning the
spins along its direction, disfavoring thus double occupancy) and
the on-site potential, which can be either attractive or
repulsive. The competition among these terms may affect the
transition temperature. To understand the effect of a magnetic
field on the critical region, in Figs. \ref{pha_diam} we plot the
phase diagram at constant $U$ for several value of $h$.
In Figs. \ref{pha_diam} we consider the full range of variation of
the particle density $0 \le n \le 2$, although it would be
sufficient, owing to the particle-hole symmetry, to explore just
the interval [0, 1].

%

\begin{figure}[t]
\centerline{\includegraphics[scale=0.31]{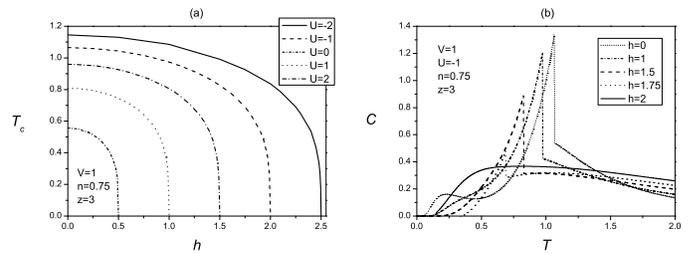}}
\caption{\label{tcvsh} (a) The critical temperature $T_c$ as a
function of the magnetic field $h$ at $n=0.75$ for $V=1$ and
$z=3$, and several values of $U$. (b) The specific heat as a
function of the temperature at $n=0.75$, for $V=1$, $U=-1$, and
$z=3$, and several values of $h$.}
\end{figure}

In the plane $(n,T)$, for attractive $U$, the CO phase is observed
in a interval $\Delta n$ which varies with the magnetic field. As
it is shown in Fig. \ref{pha_diam}a, at $h=0$ and in the limit $T
\to 0$, a complete CO state is established in the region $2/z \le
n \le 2(z-1)/z$. $\Delta n$ first increases with $T$, then
decreases vanishing at $n=1$, where the maximum critical
temperature is reached; a reentrant behavior characterizes this
region.
Upon turning on a finite magnetic field, $\Delta n$ increases by
increasing $h$ and the reentrant behavior is lost. For $h \ge
\vert U \vert$, a CO state is established in the region $1/z \le n
\le (2z-1)/z$ (in the limit $T \to 0$). The phase diagram still
presents a single lobe structure centered at $n=1$, although the
height of the lobe has decreased. There exists a critical value of
the magnetic field $h_T=zV/2-U/2$ above which one observes the
formation of three separated lobes centered around $n=0.5$, $n=1$
and $n=1.5$, respectively. The transition to the CO phase is
suppressed in the range $0.75<n<0.85$ and $1.15<n<1.25$ (for
$z=3$). By further increasing $h$, the central lobe shrinks and
eventually disappears. As one can infer from Fig. \ref{pha_diam}b,
the picture is very similar for repulsive on-site interaction. The
phase diagram presents a single lobe structure centered at $n =
1$: a CO phase is observed below the critical temperature in the
range $1/z<n<(2z-1)/z$, which does not depend on $h$. At the same
value of the magnetic field $h_T=zV/2-U/2$, the CO phase is
observed inside the three separated lobes as for the case $U<0$.
When plotted as a function of the magnetic field, the critical
temperature shows a decreasing behavior, which is more pronounced
in the neighborhood of $n=1$. Moreover, as one can also notice
from Figs. \ref{pha_diam}, according to the value of the particle
density, two situation can occur: either the transition
temperature is finite for all values of $h$ or it vanishes at same
critical value. In Fig. \ref{tcvsh}a, we plot the transition
temperature $T_c$ as a function of the magnetic field at $n=0.75$,
$z=3$ and for different values of the on-site potential. The
transition temperature decreases by increasing $U$ and vanishes at
$h_T=zV/2-U/2$. For $U=3V$, $h_T=0$: a strong on-site repulsion
inhibits the CO phase, as already noticed in Ref.
\cite{mancinis09}.
The study of the specific heat also enlightens the influence of
the magnetic field on the thermodynamic behavior of the system.
The specific heat is given by $C=dE/ dT$, where the internal
energy $E$ can be computed as the thermal average of the
Hamiltonian \eqref{hamilt}. As an example of the characteristic
behavior of the specific heat by varying $h$, in Fig. \ref{tcvsh}b
we plot $C$ as a function of the temperature at $U=-1$, $z=3$ and
$n=0.75$, for several values of the magnetic field. For $h=0$ and
at low temperatures, the system is in a CO phase and exhibits a
phase transition at $T_c$ to a homogeneous phase. The specific
heat exhibits a peak at $T_1=T_c$ - due to the phase transition -
and another peak  $T_2$ at low temperatures which vanishes as $h$
increases. For finite magnetic fields, the position of the peak
$T_1$ and the relative height decrease by augmenting $h$. For
$h=h_T$, the transition is suppressed and, correspondingly, the
peak $T_1$ disappears.

The competition among the magnetic field and the on-site and
intersite potentials gives rise to the formation of plateaus in
the magnetization curves. By increasing the magnetic field, one
observes plateaus whose starting points depend on the particle
density, as well as on the on-site potential: one identifies two
critical values of the magnetic field. The nonzero magnetization
can either begin from $h=0$ or from a finite field. $h_c$ denotes
the starting point of a nonzero magnetization, whereas $h_s$
denotes the value of the magnetic field when it reaches
saturation.
\begin{figure}[t]
\centerline{\includegraphics[scale=0.31]{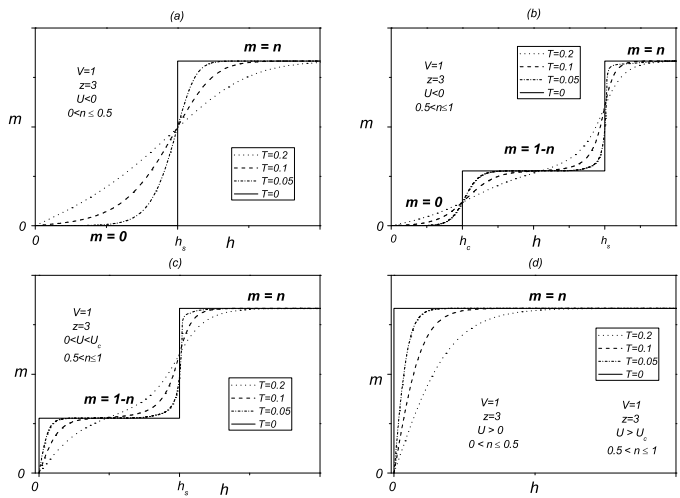}}
\caption{\label{magnet} The magnetization $m$ as a function of the
magnetic field $h$ at $T=0$.}
\end{figure}
The results for the magnetization $m(h)$ are shown in Fig.
\ref{magnet}, where $U_c$ is the critical value of the on-site
potential separating the different observed behaviors; for $z=3$,
one finds $U_c \approx 2.8$. Making  reference to Fig.
\ref{magnet} for the different regions of $n$ and $U$, one has:
$h_c=h_s=\vert U \vert /2$ (Fig. \ref{magnet}a), $h_c=\vert U
\vert /2$ and $h_s=zV/2+\vert U \vert /2$ (Fig. \ref{magnet}b),
$h_c=0$ and $h_s=zV/2-U /2$ (Fig. \ref{magnet}c), and $h_c=h_s=0$
(Fig. \ref{magnet}d). The so-called metamagnetic behavior is
clearly seen: at low temperatures the magnetization begins to show
a typical $S$-shape which becomes more pronounced by further
decreasing the temperature. At $T=0$ one, two or three plateaus
($m=0$, $m=1-n$ and $m=n$) are observed, according to the values
of the external parameters. These results are similar to the ones
obtained in one-dimensional AL-EHM \cite{mancini_epjb09} and
spin-1 antiferromagnetic Ising chain with single-ion anisotropy
\cite{chen03,mancini08}.

\begin{figure}[h]
\vspace{1cm}
 \centerline{\includegraphics[scale=0.31]{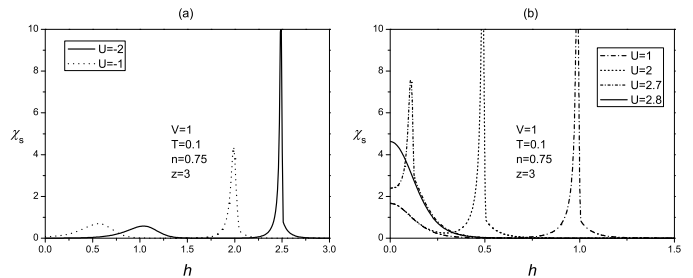}}
\caption{\label{fig8} (a) The spin susceptibility as a function of
the temperature for $V=1$, $n=0.75$, and (a) attractive  (b)
repulsive on-site interaction.}
\end{figure}

In Figs. \ref{fig8}a-b we plot the spin susceptibility as a
function of the magnetic field at $T=0.1$ for several values of
$U$ (both attractive and repulsive), $z=3$ and for $n=0.75$. In
the limit $T \to 0$, the spin susceptibility diverges in
correspondence of the values $h_{crit}$ at which the system moves
from one magnetization plateau to the other. For low values of the
magnetic field and attractive on-site interactions - corresponding
to Fig. \ref{fig8}a - the spin susceptibility vanishes at low
temperatures for all values of the filling: in the limit $T \to 0$
all electrons are paired and no alignment of the spin is possible.
By increasing $h$, the magnetic excitations break some of the
doublons inducing a finite magnetization: $\chi_s$ has a peak,
then decreases, the system having entered the successive magnetic
plateau. If $0.5<n<1$ then another peak is observed, corresponding
to the second jump of the magnetization when $h$ reaches the
saturated value $h_s$.
On the other hand, for repulsive on-site
interactions, a very small magnetic field induces a finite
magnetization (with the exception of $n=1$ when $U<U_c$): $\chi_s$
has a maximum at $h=0$ and then decreases by augmenting $h$,
unless another transition line is encountered, as it happens for
$0.5<n \le 1$ and $0<U<U_c$.


\section{Concluding remarks}

The Green's function and equations of motion formalism allows one
to tackle a large class of classical fermionic and spin systems,
providing a general formulation for any dimension and any
underlying lattice \cite{Mancini05,mancini2,mancini05b,mancini08}.
In this paper we have evidenced how the use of this formalism
leads to the exact solution of the AL-EHM on the Bethe lattice in
the presence of an external magnetic field.  By considering
nearest-neighbor repulsion $V$, there is a transition temperature
below which a charge ordered phase, characterized by a different
distribution of the electrons in alternating shells, is
established for $n>1/z$. The onset of the CO phase is signalled by
the breaking of translational invariance: at $T_c$ the values of
both the particle density and  the double occupation become site
or, more properly, shell dependent \cite{mancinis09}. The charge
ordered phase is dramatically affected by the presence of a
magnetic field: the transition temperature is lessen by a finite
magnetic field.
 The CO phase shrinks by increasing
$h$, leading to the appearance of large regions where the strength
of the magnetic field prevents the ordering of the particles.
By investigating the magnetic properties of the system, we found
magnetic plateaus at low temperature. Furthermore, we identified
the values of the critical fields $h_c$ and $h_s$, defining the
beginning point of nonzero magnetization and the saturated
magnetization field, respectively.

\section*{Acknowledgements}
I thank  F. Mancini for interesting and fruitful discussions.

\end{document}